\documentclass{PoS}

\title{The Pulsar in the Crab Nebula}

\ShortTitle{The Pulsar in the Crab Nebula}

\author{\speaker{Natalia Lewandowska}%
         \thanks{The author wishes to thank the organizers of the Frontier Research in Astrophysics workshop for the invitation and the possibility to give this talk.}\\
        Astronomy Department, University of W{\"u}rzburg, Germany\\
        E-mail: \email{natalia.lewandowska@physik.uni-wuerzburg.de}}

\abstract{The Crab pulsar belongs to one of the most studied stellar objects in the sky. Since its accidental detection in 1968, its pulsed emission has been observed throughout most of the electromagnetic spectrum. Although currently one of more than 2000 known pulsars, its way of work has remained not understood making the Crab pulsar an object of continuous studies and interest. Referring to the pulsed emission of the Crab pulsar only at radio wavelengths, it reveals a diversity of different phenomena. They range from deviations of the predicted slowing down process of the pulsar with time (long time phenomena) to an irregularity of its single pulse emission (short time phenomena). Similar and different kinds of deviations are observed at other wavelengths. Consequently, the Crab pulsar provides a large diversity of different emission characteristics which have remained diffcult to interpret with a uniform theoretical approach including all observed properties.\\
Since a review of all currently examined properties of the Crab pulsar is beyond the scope of this paper, its goal is to give an overview of previous studies of the Crab pulsar predominantly at radio and $\gamma$-wavelengths with an emphasis on a possible connection to its radio single pulse emission. A discussion of a possible identification of common emission mechanisms is given.}

\FullConference{Frontier Research in Astrophysics\\
26 - 31 May, 2014\\
Mondello (Palermo), Italy}

\begin{document}

\section{Introduction}
\begin{figure*}[!ht] 
\begin{center}
\includegraphics[width=120mm,height=80mm,angle=0]{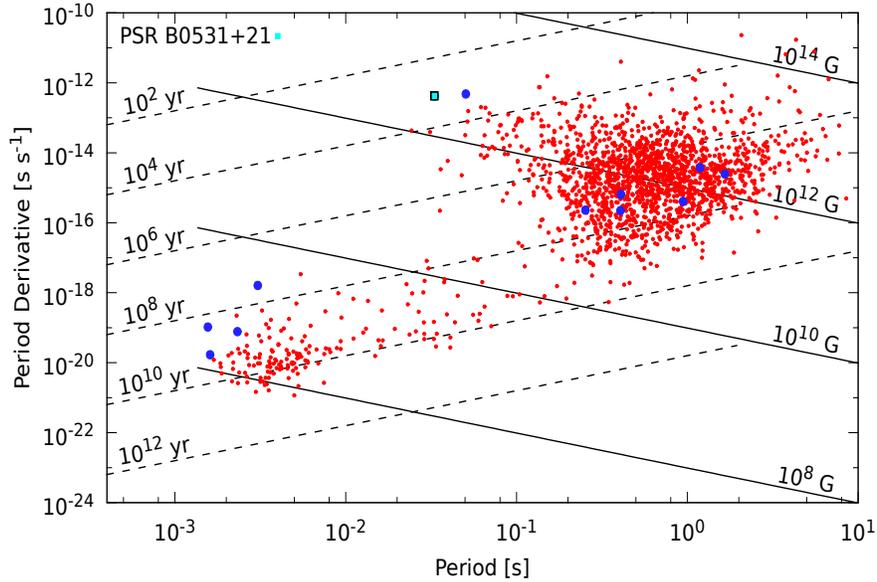}
\caption{P0-P1 diagram of all currently known pulsars. The ``island`` of points to the top right shows the group of ordinary pulsars while the objects in the lower left part of the diagram are recycled pulsars. The blue points indicate pulsars which are known to emit radio Giant Pulses (the Crab pulsar being emphasised in turquoise blue). The data for this plot was taken from the ANTF pulsar catalogue (\cite{manchester_2005}).} \label{atnf_data}
\end{center}
\end{figure*}

\begin{figure}[!ht]
\begin{center}
\includegraphics[width=120mm,height=80mm,angle=0]{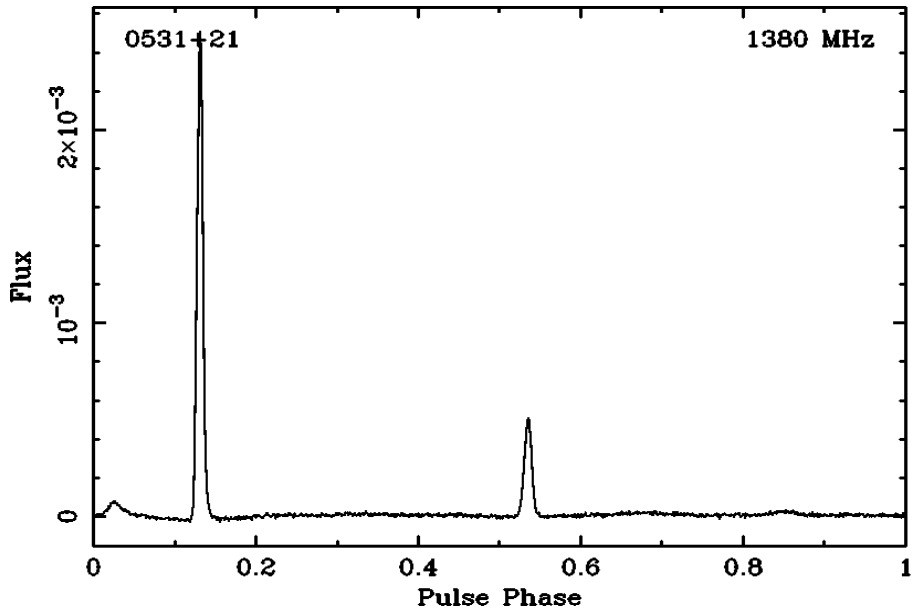}
\caption{Average profile of the Crab pulsar consisting of the Precursor around 0.5 in phase, the Main pulse at about 0.15 in phase and the Interpulse at 0.52 in phase. The data was taken with the Westerbork Synthesis Radio Telescope (WSRT). The Westerbork Synthesis Radio Telescope (WSRT) is operated by ASTRON (the Netherlands Institute for Radio Astronomy) with support from the Netherlands Foundation for Scientific Research (NWO).} \label{wsrt_observations}
\end{center}
\end{figure}

The detection of the first pulsar by Jocelyn Bell Burnell and Antony Hewish (\cite{hewish_1968}) marks the beginning of a new era of studies of neutron stars. Nowadays more than 2000 isolated pulsars are known (ANTF catalogue, \cite{manchester_2005}).
In Figure~\ref{atnf_data} the P0-P1 diagram of all currently known pulsars is shown (P0 being the rotation period and P1 its first
derivative expressing the spin down of a pulsar). It indicates the existence of several groups of pulsars 
: Ordinary pulsars (largest group in the top right corner) and millisecond pulsars
(large group in the lower left corner). Although both kinds of pulsars are rotation-powered (their rotation being their only source of
energy), their evolution history is different. While ordinary pulsars are assumed to be born in supernova explosions, millisecond pulsars are part of a binary system in which they are formed due to the accretion of matter from their companion star. Due to 
this process they spin up resulting in much smaller rotation periods and lower spin down rates than the ones from ordinary pulsars.\\
The Crab pulsar belongs to the group of ordinary pulsars and was discovered in 1968 as one of the first pulsars by accident during drift scans carried out at 112 MHz (\cite{staelin}). 
At the time of its discovery it was not clear that the detected single, bright pulses belonged to a periodically emitting source. The 
heliocentric period of the Crab pulsar was determined shortly after its discovery as 33.09144 $\pm$ 0.00001 ms (\cite{lovelace_1968}).
The location of the source was reported by \cite{lovelace_1968} to be within 10 arcminutes from the Crab Nebula center. This was later 
confirmed by further radio observations reported by \cite{comella_1969}.
Studies of the regular pulsed emission at 430 MHz revealed the existence of three components in the average profile from the Crab 
pulsar consisting of the brightest component referred to as the ``Main Pulse``, a second component called the ''Interpulse'' and an additional component emerging at earlier rotational phases than the Main pulse and therefore called the ``Precursor`` (\cite{rankin_1970}, Figure~\ref{wsrt_observations}).
Later studies even confirmed the existence of seven pulsed components (\cite{moffett_hankins_1996}) and more recent it has been  
reported  their occurrence is frequency dependent (\cite{hankins_2015}).
That the separation between the Main and the Interpulse components is increasing with time thus indicating an evolution of the magnetic field of the Crab pulsar, was detected by \cite{lyne_2013}. Besides the Crab pulsar also shows deviations with regard to the slowing down process of its rotation rate as was recently reported by \cite{lyne_2015} who analysed 45 years of data from this pulsar.\\
The fact that the Crab pulsar was not discovered by periodic pulses, but by sporadic ones, gave the impetus for a new branch of
single pulse studies. In the further part of this paper a short overview is given on this
kind of single, bright radio pulses which were later named ``Giant Pulses''.
After their detection (\cite{staelin}), further studies of these single pulses were carried out by \cite{sutton_1971} at 115, 
157 and 230 MHz. The intensity distribution of pulses from the Crab pulsar had a power law tail towards higher energies and indicated thus the possible existence of two different kinds of single pulses. 
This observation was later confirmed by \cite{argyle_gower_1972} who observed the Crab pulsar at 146 MHz and examined the distribution of 
the intensities of its single pulses. According to their measurements the intensity distribution of Crab pulsar single, bright pulses had 
the form of a power-law instead of lognormal, or even more complicated distributions which were detected in the case of regular pulses
( \cite{burke_spolaor_2012}). 
The studied single, bright pulses were later commonly referred to as Giant Pulses due to their high flux densities. 
This property was already noticed at the time of their detection (\cite{staelin}) while later studies showed that these single radio pulses well correlate with their widths (\cite{popov_stappers_2007}).
The authors did not find any Giant Pulses with widths larger than 16 $\mu$s and flux densities higher than 1000 Jy. High time resolution observations revealed the existence of 
Giant Pulses at the phases of the Main pulse with widths less than 0.4 ns and flux densities higher than 2 MJy (\cite{hankins_2007}).
Radio Giant Pulses have been observed to occur only at the phase ranges of the Main and the Interpulse (\cite{jessner_2010}).\\ 
The mentioned studies indicated that the single, radio pulses once detected by \cite{staelin} were apparently a different form of Crab 
pulsar radio emission. Since the detection of the Crab pulsar, other radio pulsars have also been found to emit this kind of single, radio
pulses as seen in Figure~\ref{atnf_data} of \cite{knight_2006}. The currently known group of giant pulse emitting pulsars consists of ordinary 
pulsars as well as millisecond pulsars. A common emission mechanism between all of them is currently not known. 
One question which is still open is if there are different sorts of giant
pulse emitters among these pulsars. Another question, which is also discussed in this report, is if its giant radio pulses can be regarded 
as the still missing link to the understanding of the mechanism behind the multiwavelength pulsed emission of the Crab pulsar.

\section{Multiwavelength Observations of the Crab Pulsar}
One specific characteristic of the Crab pulsar is that since its discovery at radio wavelengths (\cite{staelin}) it was also discovered 
at optical (\cite{cocke_1969}), X-ray (\cite{fritz_1969}), infrared (\cite{neugebauer_1969}) and $\gamma$-wavelengths (\cite{vasseur_1970}).
Unlike other pulsars, the Crab pulsar has been detected overall the electromagnetic spectrum. Its pulsed emission has so far been observed in a frequency range from 20 MHz (\cite{ellingson_2013}) up to 
96 564 195 EHz (\cite{aleksic_2012}), or from 10$^{-8}$ eV to 400 GeV.\\
A comparison of Crab pulsar light curves resulting from multiwavelength observations is shown in Figure 1 of \cite{abdo_2010}. It emphasises that the mechanism which generates the pulsed emission at radio wavelengths for instance, might be connected with the mechanisms also at other wavelengths since the pulses at different wavelengths are aligned in phase.
\\
\\
From the theoretical point of view the emission mechanisms which cause this alignment, are currently not understood as the radio emission 
was thought to develop near the polar caps of the neutron star whereas the $\gamma$-ray emission is assumed to develop at the Outer Gaps 
of the pulsar magnetosphere (see Chapter 3.6 and 3.7 in \cite{lorimer_kramer_2012}). Two different models were suggested as an explanation 
of the observed $\gamma$-ray spectra up to 400 GeV (\cite{aleksic_2012}).
The Non-Vacuum Outer Gap model proposed by \cite{hirotani_2006} is an extension of the original Outer Gap model (\cite{cheng_1986}).
It describes the existence of the very high energy emission up to 400 GeV as a consequence of the inverse Compton scattering of secondary
and tertiary electrons and positrons on infrared and ultraviolet photons.
Another approach proposed by \cite{aharonian_2012} states that the acceleration zone of particles is located outside of the light
cylinder (a cylindrical region centered on the pulsar which is corotating at the speed of light at its radius) in a zone of 20 to 50 
light cylinder radii. With this approach they are capable of reproducing the observed spectra with a cutoff at approximately 500 GeV.
Both models predict emission regions which are further away from the pulsar. Hence the question remains open, how and if the emission from
the two proposed regions of $\gamma$-rays and from the polar caps might be connected.\\
A comparison between the arrival times of optical and radio photons resulted in a delay of 255 $\pm$ 21 $\mu$s with the optical pulse leading (\cite{oosterbroek_2008}), between X-ray and radio photons with the X-ray pulse leading with regard to the radio pulse by 344 $\pm$ 40 $\mu$s (\cite{rots_2004}) and between radio and $\gamma$-ray pulses in a delay of 241 $\pm$ 21 $\mu$s with a leading $\gamma$-ray pulse (\cite{kuiper_2003}).\\
In the case of Crab pulsar Giant Pulses several correlation studies at various wavelengths were carried out (\cite{bilous_2012}, \cite{mikami_2014}, \cite{bilous_2011}, \cite{mickaliger_2012}).
Currently, only a correlation with optical photons is known (\cite{shearer_2003}). The optical flux was observed to increase by 3 \% during occurring radio Giant Pulses. This detection was independently confirmed by \cite{strader_2013} and is so far the only reported correlation. It implies an additional incoherent emission mechanism linked to the coherent radio emission which triggers Giant Pulses.\\
As already mentioned also other pulsars which emit radio Giant Pulses have been detected, the second being  
the millisecond pulsar PSR B1937+21 (\cite{backer_1982}). In this case Giant Pulses have been observed to occur at the trailing edge of the
regular emission profile in a time difference of about 55 to 70 $\mu$s after its Main and Interpulse (\cite{kinkhabwala_thorsett_2000}).
X-ray observations revealed that the X-ray pulses of this pulsar are close in phase to its radio giant 
pulses (\cite{cusumano_2003}) suggesting similar emission regions in the magnetosphere.\\
Studies of other giant pulse emitting pulsars (for an overview see \cite{knight_2006}), revealed also the 
existence of closely aligned pulsed high energy emission (meaning in this context higher than radio wavelengths) in other cases.
Extensive surveys for giant pulse emitting pulsars carried out by \cite{romani_johnston_2001} revealed the existence of radio Giant Pulses
in the case of the millisecond pulsar PSR B1821-24. Furthermore they reported the observed Giant Pulses to be located in a narrow phase
window which coincides with the hard X-ray pulses from this pulsar. This example suggests again a possible connection between radio giant 
pulses and high energy emission.\\
No common property between all giant pulse emitting pulsars is currently known. 
Though for a long time the occurrence of Giant Pulses was supposed as due to a high magnetic field at the light cylinder, this  
criterion is now critically discussed because of the detection of older ordinary pulsars which have been detected to emit single, bright pulses that qualify as Giant Pulses (\cite{ershov_2003}).
Future surveys with more sensitive instruments like the SKA will be decisive in solving the still open questions behind the radio giant 
pulse emission from the Crab pulsar as well as from other giant pulse emitting pulsars.


\end{document}